\documentclass[runningheads]{llncs}
\usepackage{graphicx}
\usepackage{url}
\usepackage{enumerate}
\usepackage{enumitem}
\usepackage[dvipsnames]{xcolor}
\usepackage{float}
\usepackage{soul}
\usepackage[super]{nth}
\usepackage{booktabs}
\usepackage{svg}
\usepackage[nocompress]{cite}

\begin{document}
\title{Design of Secure Coding Challenges for Cybersecurity Education in the Industry}
\titlerunning{Design of Secure Coding Challenges for Education in the Industry}
\author{
  Tiago Gasiba\inst{1,2}\orcidID{0000-0003-1462-6701}
  \and
  Ulrike Lechner\inst{2}\orcidID{0000-0002-4286-3184}
  \and \\
  Maria Pinto-Albuquerque\inst{3}\orcidID{0000-0002-2725-7629}
  \and \\
  Alae Zouitni\inst{4}\orcidID{0000-0002-8809-7657}
}
\authorrunning{T. Gasiba et al.}

\institute{
  Siemens AG, Munich, Germany \email{tiago.gasiba@siemens.com}
  \and
  Universität der Bundeswehr München, Munich, Germany \email{ulrike.lechner@unibw.de}
  \and
  Instituto Universitário de Lisboa (ISCTE-IUL), ISTAR-IUL, Lisboa, Portugal \email{maria.albuquerque@iscte-iul.pt}
  \and
  Universität Passau, Passau, Germany \email{zouitni.alae@gmail.com}
}

\maketitle

\begin{abstract}

According to a recent survey with more than 4000 software developers, \textit{"less than half of developers can spot security holes"}.
As a result, software products present a low-security quality expressed by vulnerabilities that can be exploited by cyber-criminals.
This lack of quality and security is particularly dangerous if the software which contains the vulnerabilities is deployed in critical infrastructures.
Serious games, and in particular, Capture-the-Flag(CTF) events, have shown promising results in improving secure coding awareness of software developers in the industry. The challenges in the CTF event, to be useful,  must be adequately designed to address the target group. This paper presents novel contributions by investigating which challenge types are adequate to improve software developers' ability to write secure code in an industrial context. We propose 1) six challenge types usable in the industry context, and 2) a structure for the CTF challenges. Our investigation also presents results on 3) how to include hints and penalties into the cyber-security challenges. We evaluated our work through a survey with security experts. While our results show that "traditional" challenge types seem to be adequate, they also reveal a new class of challenges based on code entry and interaction with an automated coach.


\end{abstract}

\section{Introduction}
\label{sec:introduction}
To improve the quality (ISO250xx~\cite{ISO250xx}) of software in terms of security, several standards such as IEC-62443-4-1~\cite{2018_62443_4_1} and ISO 27001~\cite{2013_27002} mandate the implementation of a secure software development lifecycle (S-SDLC).
Additionally, in recognition of the importance of secure code and need to develop secure products~\cite{gitlab_2019,schneier_2019_sw_devel}, several companies have joined together and formed the SAFEcode~\cite{SAFECode} alliance to promote security best practices.
Automatic tools such as Static Application Security Testing (SAST)~\cite{rodriguez2019software} can be used to automate and aid in improving code quality.
These tools scan the code basis for existing vulnerabilities, which must be fixed by software developers.
However, previous research shows that this is not enough~\cite{oyetoyan2018myths}: the reliability of such tools is still not good enough, and they cannot automatically fix the code - this is done by software developers who must also be trained in secure software development.

One of the methods currently being investigated and that is showing promising results are training methods based serious games of the type Capture-the-Flag (CTF).
The concept of these kinds of games was originally developed in the pen-testing community.
Several such games are continually being deployed around the world~\cite{CTFTIME} nowadays by universities, companies, and even groups of individuals.
However, most of the existing CTFs are not geared towards software developers in the industry.
Gasiba et al.~\cite{gasiba_re19} have recently shown that, in order to raise awareness on secure coding in the industry, the game design must address the specific requirements of its target audience.

Typically CTFs can be categorized as follows: 1) \textit{Attack-Only}, 2) \textit{Attack-and-Defend} and 3) \textit{Defend Only}.
The participants of these CTFs are generally split into two categories: Red Team (attackers) and Blue Team (defenders).
In \textit{Attack-Only} Red team players try to exploit several systems to gain access and control.
In \textit{Attack-and-Defend} competitions, the Red team players attack systems that are being hardened and protected by blue team members.
Finally, in \textit{Defend Only} CTFs, the players answer questions on cybersecurity for points or configure and harden systems to be resilient to simulated attacks.

To address the needs of the industry and to better adapt to the players, Gasiba et al.~\cite{gasiba_re19} have proposed a defensive CTF approach and also outlined the requirements for the design of the defensive challenges.
A proper design of challenge types based on these requirements is especially important in an industrial setting, as shown by an experiment by Barela et al.~\cite{2019_barela}, where the type of the challenge (based on comics) was seen to be inadequate for CTFs in the industry.

Therefore, in this paper, we extend previous work by addressing the question of which types of defensive challenges are suitable for software developers.
In particular, we are interested in the 1) structure of the said challenges and also on 2) which types of challenges can be used in a CTF-like competition to raise awareness of software developers in the industry.
Our work is based on surveys administered through interviews with expert security trainers from the industry. The main contributions of this work are the following:
\begin{itemize}
    \item design of defensive challenges for CTFs in the industry which aim at raising awareness on secure coding and secure coding guidelines
    \item definition of a challenge structure for industrial CTFs,
    \item definition of six different challenge types for industrial CTFs, and
    \item insight into different options on how to include hints and penalties in industrial CTFs.
\end{itemize}

We hope that this work can be used by designers of serious game and quality engineers as a guideline on how to design defensive challenges for CTFs aimed at raising awareness on secure coding on software developers in the industry.

In section~\ref{sec:related_work}, we present previous work related to our research. 
Section~\ref{sec:approach} discusses our approach to the design of the defensive challenges.
The results of our study are presented in Section~\ref{sec:results}.
This section also presents a critical discussion on the obtained results, presents our main contribution to practical scenarios for possible games, and briefly discusses the threats to the validity of our findings.
Finally, section~\ref{sec:conclusions} summarizes our work and briefly discusses possible next steps.
\section{Related Work}
\label{sec:related_work}
In~\cite{Graziotin2018}, Graziotin et al. have shown that \textit{happy developers are better coders}, i.e., produce higher quality code and software.
Davis et al. in~\cite{2014_fun_and_future_of_ctf} show that CTF players experience fun during game play.
Furthermore, Woody et al.~\cite{woody2015predicting} argue that \textit{software vulnerabilities are quality defects}.
Since fun and happiness are inter-related~\cite{tews2019does}, these facts can be seen as a motivator to use Capture-the-Flag (CTF)-base serious games~\cite{2016_Doerner_Serious_Games} to raise awareness~\cite{2014_Benenson_Defining_Security_Awareness} on the topic of secure coding for software developers in the industry, in order to improve code quality.

In~\cite{Mirkovic2018Class} Mirkovic et al. introduced classroom CTF exercises as a form of cybersecurity education in academia.
Their results show that the students that participated in this kind of event have enjoyed the training and have shown increased interest, attention, and focus towards cybersecurity topics.
Additionally, in their study, Gonzalez et al.~\cite{Gonzalez2017} shown similar results and state that cybersecurity training through serious games improves the students' education and skills, and has a positive impact on attracting students to cybersecurity field.
They conclude that this kind of training can reduce the shortage of professionals in the field of cybersecurity.
Several additional studies~\cite{2014_fun_and_future_of_ctf,Beuran2016TowardsTraining, Tomomi_Aoyama2017} also show the positive benefits of CTF in students' attention and performance.

However, using CTFs as a tool to raise cybersecurity awareness comes with different obstacles.
In~\cite{gasiba_re19}, Gasiba et al. elicit requirements for designing CTF challenges geared towards software developers in the industry and show that these CTF challenges should focus on the defensive perspective.
Chung et al.~\cite{Chuang2014} also evaluated different aspects related to CTFs and concluded two important issues related to CTFs: the challenge difficulty level and suitability the target audience.

In our work, we are interested in designing high-quality defensive CTF challenges for software developers in the industry that address the topic of secure coding guidelines~\cite{CERT-SEI} (SCG) and secure software development best practices~\cite{owaspT10} (SDBP).
However, most of the currently existing work focuses on academia, where the target group is composed of current or future security experts, or pen-testers.
Furthermore, most existing studies also focus on the offensive perspective and do not address the topic of SCG, and SDBP.
As such, this study is driven by both the need to raise awareness on secure coding~\cite{Acar2017,Nance2012,Yang2016}, and by the lack of design of defensive CTF challenge geared towards software developers in the industry.
The research method used in this work is based on semi-structured interviews~\cite{Lisa2008} and survey best practices as described by Grooves et al.~\cite{Groves2009} and Seaman et al.~\cite{Seaman1999}.
The design of the serious games is based on~\cite{2016_Doerner_Serious_Games}.
\section{Approach to Challenge Design}
\label{sec:approach}
\begin{figure}[ht]
    \centering
    \includegraphics[width=.8\columnwidth]{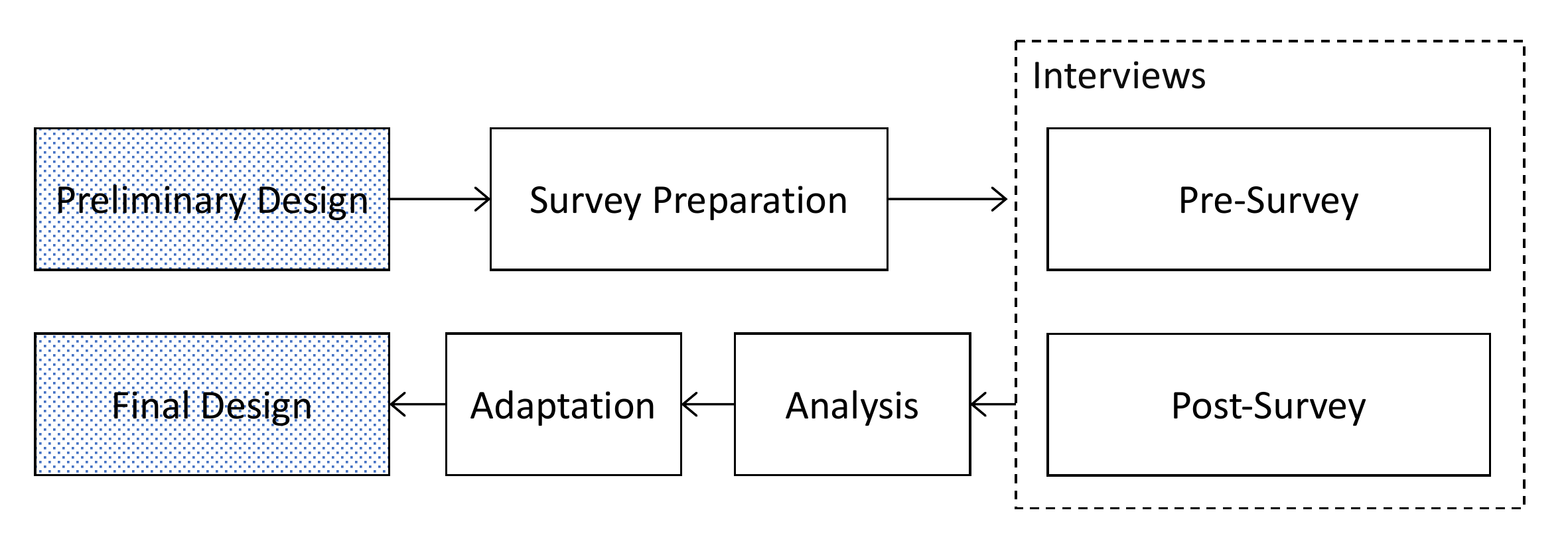}
    \caption{Study Approach}
    \label{fig:approach}
\end{figure}

In order to design defensive challenges for industrial CTFs, the authors have decided to focus on two different aspects: the challenge structure (CS) and the challenge type (CT).
The content of the challenge (e.g. questions or example of software vulnerability), which are not the focus of the current work, can be derived from existing SCG~\cite{CERT-SEI} and SDBP~\cite{owaspT10}.
The challenge structure reflects the mechanics of the challenge, i.e., how it is supposed to be deployed and how it should work.
The challenge type specifies the different ways that the challenge can be presented to a participant.
Figure~\ref{fig:approach} shows the steps that we have followed in our approach to design the defensive CTF challenges for an industrial context.

In the first step, we have created a preliminary design, containing a proposed CS and different CTs.
For this, we conducted several informal discussions with one security expert. Additionally, based on our experience with past Capture-the-Flag events, we concluded the preliminary design of challenge structure and challenge types.
In the next step, we created a two-phase survey~\cite{Groves2009,Seaman1999}.
The goal of the survey was to gather feedback and opinions, in a structured way, on the preliminary challenge structure and challenge types.
It was used to facilitate the semi-structured interviews with several security experts.
The interviews, carried out in the following step, were realized in face-to-face meetings.
The meetings consisted of three parts: pre-survey, post-survey, and informal discussions.
After the interviews, the collected feedback was transferred to digital form and was analyzed.
The analysis step aims at understanding the joint agreement on the different suggested improvements by the security experts.
The commonly suggested improvements were then used to adapt and change the preliminary design, which resulted in the final challenge design.

In the following sub-sections, we present details on the different phases of our approach.
The results of the analysis, adaptation, and also the final challenge design will be presented in section~\ref{sec:results}.

\subsection{Preliminary Design}
In the preliminary design, the authors conducted several informal discussions with a security expert which is also a trainer of secure coding in the industry.
The security expert has more than 10 years of experience in the industry and has also knowledge and had previously participated in Capture-the-Flag events.
Based on the experience of the security expert and also on the experience of the authors, preliminary design was derived.

\subsection{Survey Preparation}
In order to prepare for the interviews with security experts, a two-part survey~\cite{krosnick2018questionnaire} was developed by the authors.
The developed survey underwent three reviews by three different cybersecurity experts: one holding a master of science in computer science and two holding a Ph.D. in IT security, whereby one is additionally a university lecturer in cybersecurity.
The main goal of the pre-survey was to understand what types of challenges do experienced industry security experts find suitable for CTF-based awareness training.
The post-survey' primary goal is to understand the level of agreement with the different preliminary challenges types.
The pre-survey was conducted at the beginning of the meeting, before presenting the preliminary design. The post-survey was conducted after presenting the preliminary design.
This split allowed the participants to think and reflect on their answers from the pre-survey and be prepared and more open-minded for the discussions on the post-survey.
Splitting the survey into two parts was done in order to guarantee unbiased feedback collection from the security experts during the pre-survey.
Both the pre-survey and post-survey asked the participants - \textit{if they were to design a CTF challenge about secure coding for software developers in the industry, what kind of challenge structure and type would they use?}.
The post-survey additionally asked questions on the preliminary design, in particular on \textit{what would the participant change, add or remove to the presented preliminary design}, \textit{what other challenge types would they additionally consider} and also on the expert opinion on \textit{how to use penalties and hints in the challenges}.
In total, the pre-survey consisted of 16 questions, of which 12 were multiple-choice, three were based on a Likert scale, and 1 was an open-ended question \cite{SMITH2000}.
The post-survey consisted of 11 questions, whereby 5 were feedback questions based on a Likert scale, and 6 were open-ended questions.

\subsection{Interviews}
For the interview, the authors engaged $20$ security experts with an average of 4 years of experience in the industry (minimum one year and maximum of 12 years).
The experts were selected based on their experience, position, and background in the company - engaged in several consulting projects as a cyber-security expert.
A large part of the participants were also trainers themselves of different topics on cybersecurity.
The selected participants were all familiar with CTF competitions.
Half of the experts hold a Ph.D. degree in computer science or equivalent, and the remaining half holds a master of science in computer science or equivalent.
The face-to-face interviews lasted for one hour and were carried out between the \nth{1} of October $2019$ and the \nth{16}.
During the interview, the first $20$ minutes were dedicated to the pre-survey.
Afterward, the preliminary design of CTF challenges was presented to the participants.
The remaining $30$ minutes were then spent on the post-survey, open-ended discussions and finished with $10$ minutes of informal discussions on the results.

\subsection{Analysis}
In this stage, we gathered all the collected data from the pre-surveys, post-surveys, and informal discussions.
The results using a Likert scale were analyzed using standard statistical methods.
Due to its nature, the open-ended questions and the informal discussions need to be coded~\cite{Seaman1999}.
In order to guarantee the quality of this step, the transcripts were given to three security experts who were asked to perform the coding step manually.
We have opted for a manual procedure rather than automated to ensure high quality, as automated coding has been previously shown not to achieve high accuracy~\cite{Schonlau_Couper_2016}.
The coding outcome of each expert was then collected and discussed together.
Similarities and differences were then systematically addressed, and the final coding was derived by mutual agreement between the three experts.

\subsection{Adaptation and Final Design}
The last step consisted of using the feedback from the previous step to adapt and change the preliminary design accordingly.
Only the proposed changes that were agreed by the majority of the participants (i.e., more than $2/3$ after coding or 80\% of participants) were considered for the final design.
In section~\ref{sec:results}, the final challenge design, including challenge structure and derived challenge types, will be presented in more detail.
\section{Analysis and Results}
\label{sec:results}
In this section, we describe the results from the two-part survey interview, as outlined in the previous section.
We present the final challenge structure and types, which take into consideration the feedback provided by all the security experts.
Finally, we summarize the main contributions and briefly discuss the threat to the validity of our work.

\subsection{Preliminary Design}
As a result of the informal discussions with the security expert, the challenge structure was defined in two rounds: \textit{round 1}: main challenge and \textit{round 2}: presentation of secure coding guideline related to the challenge. No further details will be given for the initial design, as this was changed after the interview with the security experts, as shown in the next sub-sections. Section~\ref{sub:final_challenge_structure} details the final challenge structure.
The derived initial challenge types were the following: Single Choice Question (CSQ), Multiple Choice Question (MCQ), Text Entry Challenge (TEC), Code Snippet Challenge (CSC), and Code Entry Challenge (CEC).
Table~\ref{tab:question:type} shows a summary of the challenge types. Further details are given in section~\ref{sub:final_challenge_design}, together with the final design.

\subsection{Pre-Survey Results}
Pre-survey results showed that the majority ($55\%$) of the participants thought an adequate type of challenge would be of type \textit{question and answer}, without specifying what they mean.
Additionally, $85\%$ answered that \textit{some form of challenge involving coding} would be adequate, since the challenges should be based on SCG.
However, some participants replied that \textit{friendly hacking} exercises would also be a good exercise - this was discarded as these types of challenges are not defensive.
One participant mentioned that an appropriate challenge would involve \textit{fixing a problem on a vulnerable code snippet}.

{\small 
\vspace*{1em}
\begin{table}[http]
    \centering
    \renewcommand{\arraystretch}{0.8}
    \begin{tabular}{|l|l|}
      \hline
          \multicolumn{1}{|c|}{\textbf{Question}} &
          \multicolumn{1}{c|}{\textbf{Pre-survey Results}} \\
      \hline
          When would you add the hints? &
          \begin{tabular}[c]{@{}l@{}}
              (30\%) For all challenges\\
              (50\%) For difficult challenges\\ 
              (20\%) No opinion
          \end{tabular} \\
      \hline
          How to design the hints? &
          \begin{tabular}[c]{@{}l@{}}
              (50\%) Giving details on the answer\\
              (75\%) Disclosing important concept\\
              (70\%) Include an external reference
          \end{tabular} \\
      \hline
          When would you add penalties? & 
          \begin{tabular}[c]{@{}l@{}}
              (60\%) Agree to introduce penalties \\
              ~~~~~~(35\%) Retrying the same challenge \\
              ~~~~~~(65\%) When using a hint \\
              (30\%) Disagree to introduce penalties \\
              (10\%) No opinion
          \end{tabular} \\
      \hline
    \end{tabular}
    \caption{Coding Results On Hints and Penalties}
    \label{tab:pre-survey:hintsAndPenalties}
\end{table}
}

Table~\ref{tab:pre-survey:hintsAndPenalties} shows a summary of the agreement level of the participants towards questions asked during the pre-survey related the hints and penalties.
The usage of hints was backed by $80\%$ of the survey participants, for difficult challenges ($50\%$) or all questions ($30\%$).
The hints should include details on how to solve the questions ($50\%$) and point-out the secure coding concept behind the challenge ($75\%$).
The majority of the participants agreed that adding an external reference (e.g. link to an article on the web) is an appropriate way to design hints for challenges.
Half of the participants agree that hints should disclose the essential concept behind the challenge, e.g., on which secure coding guideline the challenge is based.
Only $75\%$ of the participants agree that giving targeted hints (e.g., disclosing an important concept) is a good idea.
During the informal discussions, several participants mentioned that the goal of the hints should be to make sure that the CTF players are learning secure coding concepts during the game.
The hints should also be designed in order to lower player frustration and maximize the learn-effect.
In particular, the types of hint should be precise and to the point, as industry players have a limited time to play the game.

In terms of penalty-points, $60\%$ agreed to introduce them, $30\%$ disagreed, and $10\%$ had no opinion.
The ones that agreed to introduce penalty points, $65\%$ agreed that using hints should be penalized, and the remaining $35\%$ agreed that retrying a challenge should be penalized.
During the informal discussions, the survey participants mentioned that the intention to add penalties should be to motivate the player to find solutions by him/herself and not to rely on hints.
Furthermore, the penalties should be small to lower the frustration level while maximizing the learning effect of the CTF players.

\subsection{Post-Survey Results}
In the post-survey, the participants were shown all the derived challenge types and were asked to rate their agreement on the suitability for a CTF-like event with software developers in an industrial setting on a Likert scale.
Table~\ref{tab:level_agr} shows the results of the post-survey for the five different challenge types.
We use the standard mapping of the Likert scale as follows: from 1-{\it strongly disagree} to 5-{\it strongly agree}.

\begin{table}[ht]
\centering
\renewcommand{\arraystretch}{0.8}
\begin{tabular}{l|l|l|l|l|l|}
\cline{2-6}
\multicolumn{1}{c|}{}                         & \multicolumn{1}{c|}{\textbf{SCQ}} & \multicolumn{1}{c|}{\textbf{MCQ}} & \multicolumn{1}{c|}{\textbf{TEC}} & \multicolumn{1}{c|}{\textbf{CSC}} & \multicolumn{1}{c|}{\textbf{CEC}} \\ \hline
\multicolumn{1}{|l|}{\textit{Average}}        & 3.95                              & 3.80                              & 3.15                              & 4.30                              & 4.30                              \\ \hline
\multicolumn{1}{|l|}{\textit{Std. Deviation}} & 0.76                              & 1.00                              & 1.04                              & 1.26                              & 0.92                              \\ \hline
\end{tabular}
\caption{Average Agreement Level}
\label{tab:level_agr}
\end{table}

The derived ranks of the preferred challenge types are the following (from highest agreement to lowest agreement): 1) Code-Entry-Challenge, 2) Code-Snippet Challenge, 3) Single-Choice Question, 4) Multiple-Choice Question and 5) Text-Entry Challenge.
Although CSC and CEC have the same average agreement level, CSC has a higher standard deviation (i.e., higher uncertainty) than CEC; therefore, we have placed CEC in first place in the rank.

When the participants were asked about ideas for additional challenge types, $80\%$ had {\it no new idea}, $15\%$ answered yes, {\it they had an additional idea} and $5\%$ had {\it no opinion}.
The additional collected ideas were the following: a) \textit{"something dynamic and fun"}, b) \textit{"associating left and right lists"} (ASL) and c)\textit{"modify code that has one vulnerability"}.
The contribution (a) and (c) could not be mapped into an existing challenge type, nor could a new challenge type be discerned. However, (b) resulted in a new challenge type.

The participants were also asked what could be added to the existing challenges.
The following additional points were collected with this question:
\begin{itemize}
    \item Provide explanation at the end of the challenge, together with the flag
    \item Add explanations on multi-stage challenges
    \item Ask which coding guideline is not being followed in a code snippet
    \item Randomize the answers and randomize of the solutions
    \item Do not forget about the fun aspect when designing the challenge
\end{itemize}

These additional points were also used to improve the final challenge structure, as shown in the next sub-section.

\subsection{Final Challenge Structure}
\label{sub:final_challenge_structure}
\begin{figure}[ht]
    \centering
    \begin{minipage}{.95\textwidth}
        \centering
        \includegraphics[width=.99\textwidth]{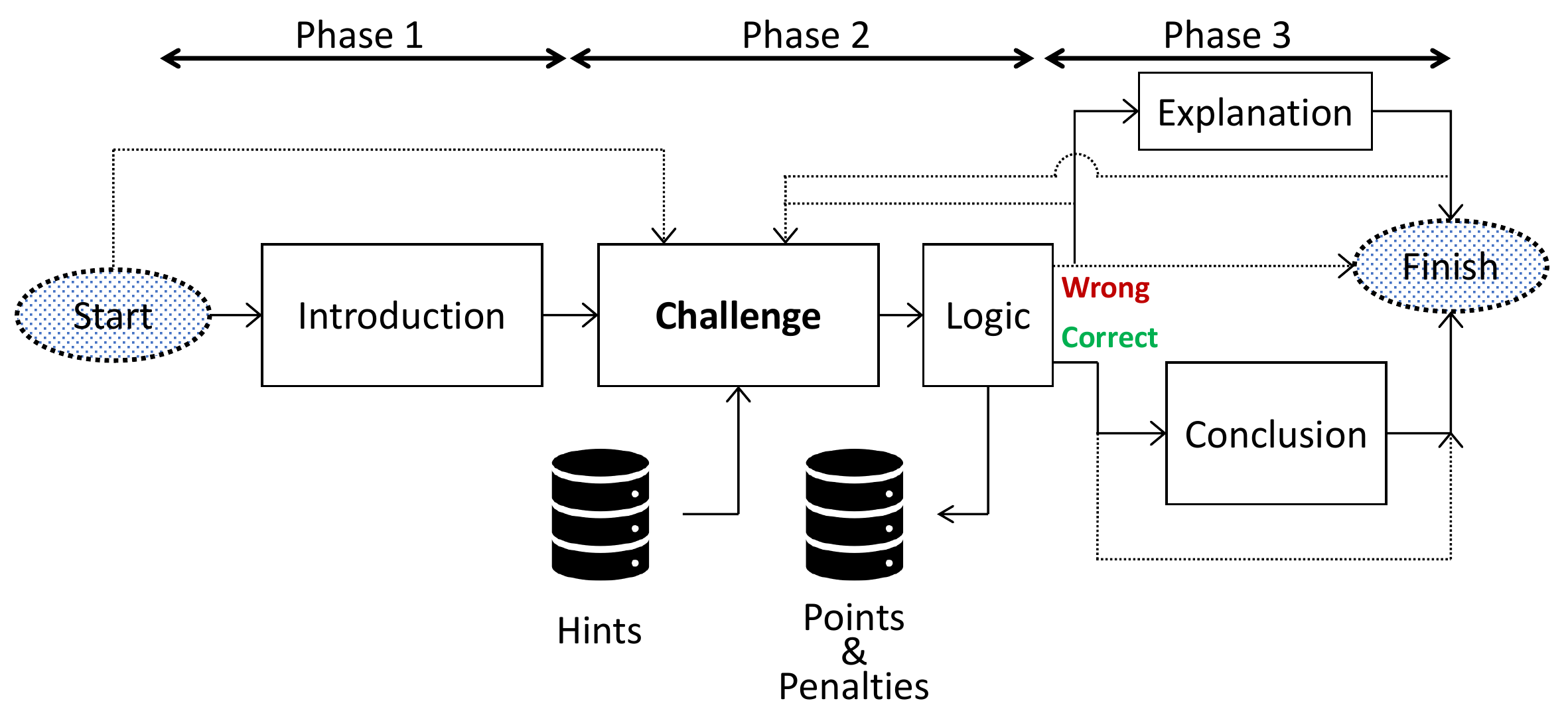}
        \caption{Challenge Structure}
        \label{fig:chal_structure}
    \end{minipage}
\end{figure}

The final challenge structure (CS) contains three phases consisting of four stages: \textit{introduction} (phase 1), \textit{challenge} and \textit{logic} (phase 2) and \textit{conclusion} (phase 3), as shown in Figure~\ref{fig:chal_structure}.
In the \textit{introduction} stage (phase 1), a topic related to secure coding is introduced, which is helpful to solve and frame the challenge (e.g., secure coding guideline or previously related cybersecurity incident).
This optional stage and can include a single-choice or multiple-choice question before proceeding to the next phase.
In the second phase, the \textit{challenge} stage contains the main CT according to a given challenge type, as presented in section~\ref{sub:final_challenge_design}.
In this stage, several hints can be given to the player depending on several factors, e.g., time taken by the player to solve the challenge or the previous number of attempts to solve the challenge.
The \textit{logic} stage is responsible for evaluating the solution to the challenge provided by the player and determining if it is correct (acceptable) or wrong (not acceptable).
According to the analysis of the answer provided by the player, points or penalties might be awarded.

The third phase depends on the result of the logic stage.
In case the player's answer was wrong, the following four options can occur: return to the challenge stage, give some explanation why the solution is wrong and return to the challenge stage, proceed to the finish or give an explanation why the solution is wrong and proceed to the finish.
In case the player's solution was correct, the following two options can occur: give some concluding remarks (with an optional additional question) and then proceed to the finish or proceed to the finish.
If a correct solution is achieved at the finish state, then a flag is presented to the player (according to the CTF rules).
In the conclusion stage, additional useful information can be given to the player, e.g., an explanation of secure coding guidelines related to the challenge, the importance of the challenge in the industry context, for example, through lessons learned from past incidents or vulnerabilities.

\subsection{Final Challenge Types}
\label{sub:final_challenge_design}
Table~\ref{tab:question:type} shows the final six derived challenge types.
In single-choice questions (SCQ), the participant is asked a question, and only one of the possible answers is the correct solution.
In multiple-choice questions (MCQ), the correct solution must include more than one different answers.
In text-entry questions (TEQ), the participant needs to type in the solution as text - this can be achieved, for example, by completing or writing a full sentence as the answer to the challenge.
Code-snippet challenge (CSC) presents a piece of code to the participant and lets the participant select lines of code containing vulnerabilities or select changes to the code that would avoid vulnerabilities (i.e., respect SCG and SDBP).
In code-entry challenges (CEC), the participant is given vulnerable code that needs to be changed or rewritten to eliminate the vulnerability by complying with SCG and SDBP.
In associate left-right challenges, the participant needs to associate items in a list on the left to items in a list on the right.

\begin{figure}[H]
    \centering
    \begin{minipage}{.45\textwidth}
        \centering
        \includegraphics[width=.99\textwidth]{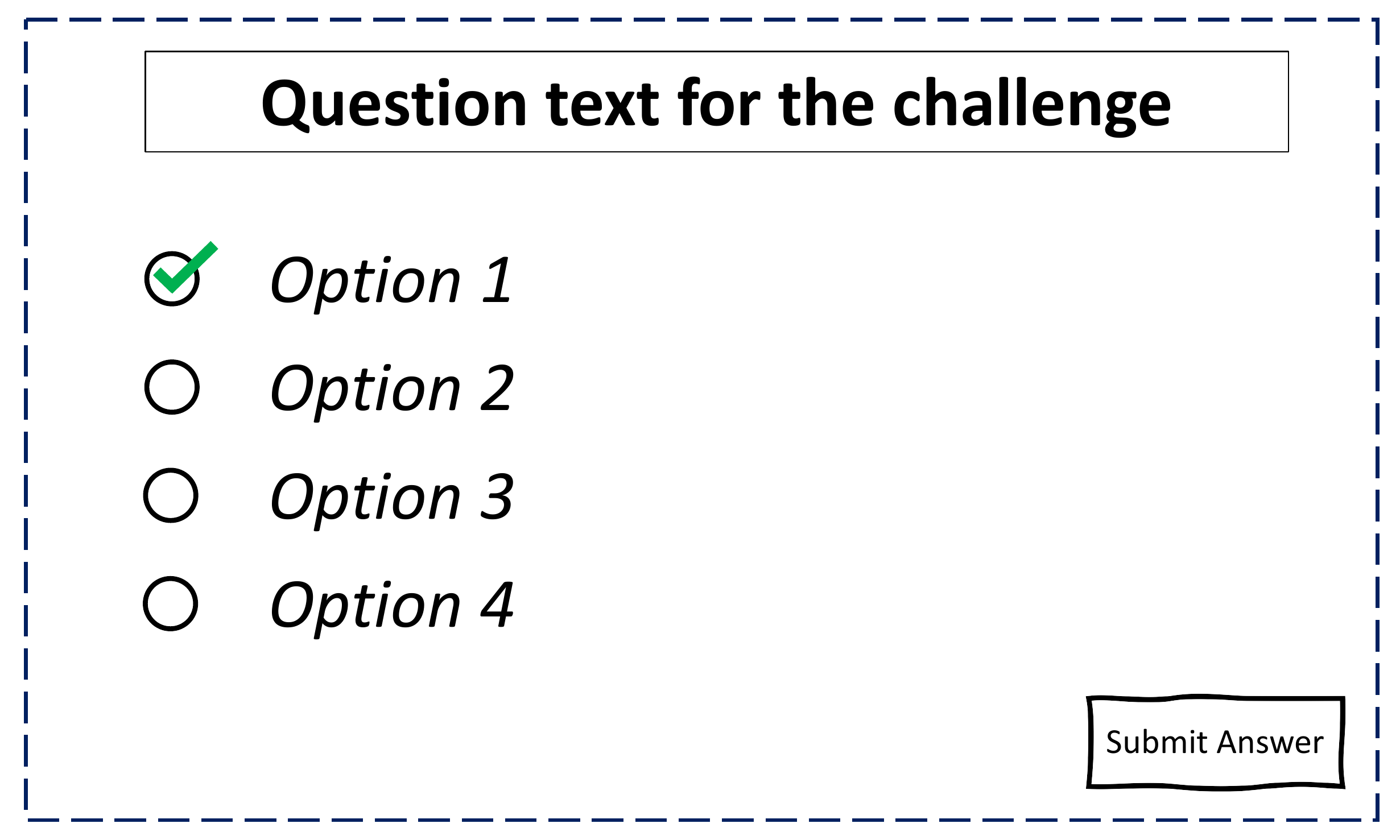}
        \vspace*{-2em}
        \caption{Single-Choice Question}
        \label{fig:generic_scq}
    \end{minipage}
    \begin{minipage}{.45\textwidth}
        \centering
        \includegraphics[width=.99\linewidth]{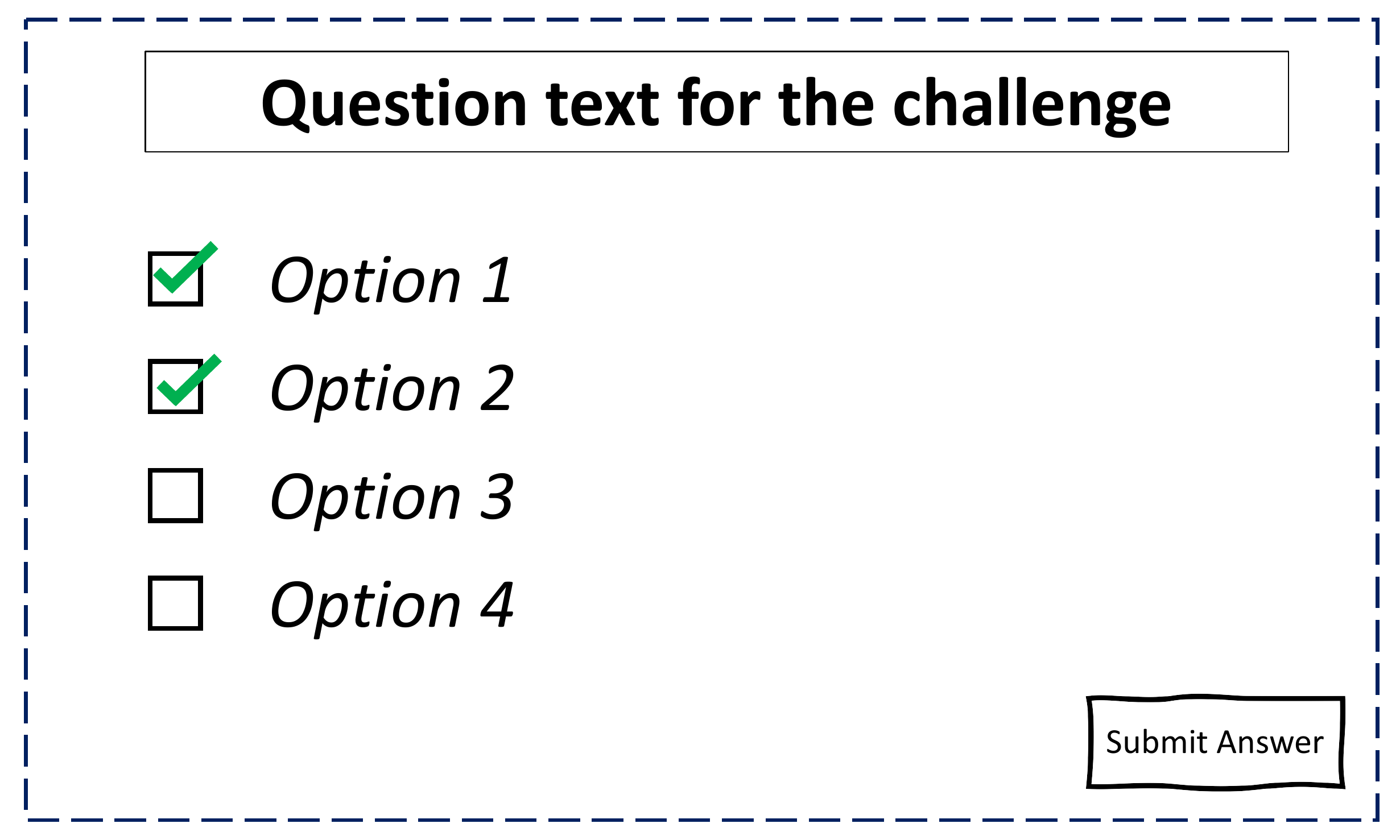}
        \vspace*{-2em}
        \caption{Multiple-Choice Question}
        \label{fig:generic_mcq}
    \end{minipage}

    \vspace*{1em}
    \begin{minipage}{.45\textwidth}
        \centering
        \includegraphics[width=.99\textwidth]{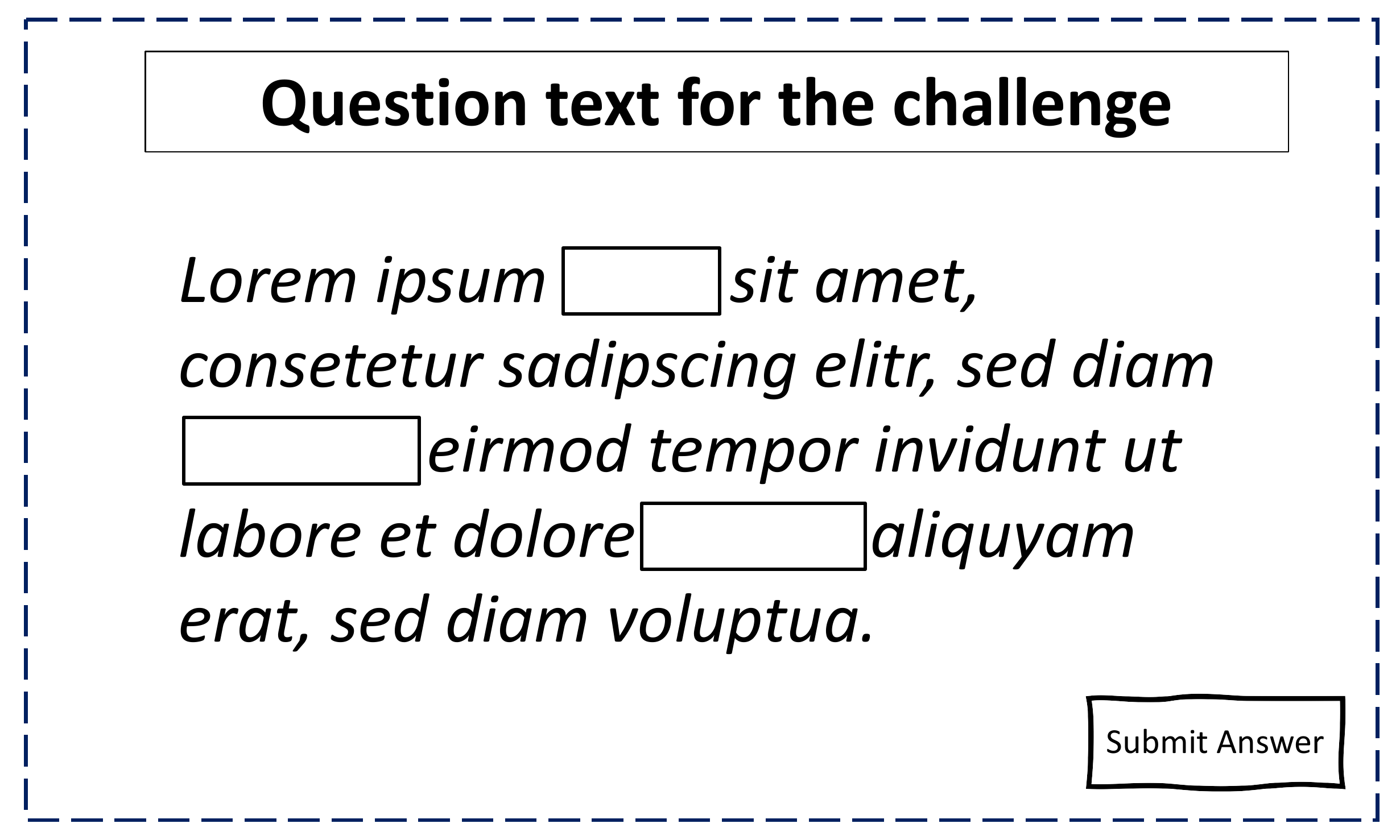}
        \vspace*{-2em}
        \caption{Text-Entry Question}
        \label{fig:generic_teq}
    \end{minipage}
    \begin{minipage}{.45\textwidth}
        \centering
        \includegraphics[width=.99\linewidth]{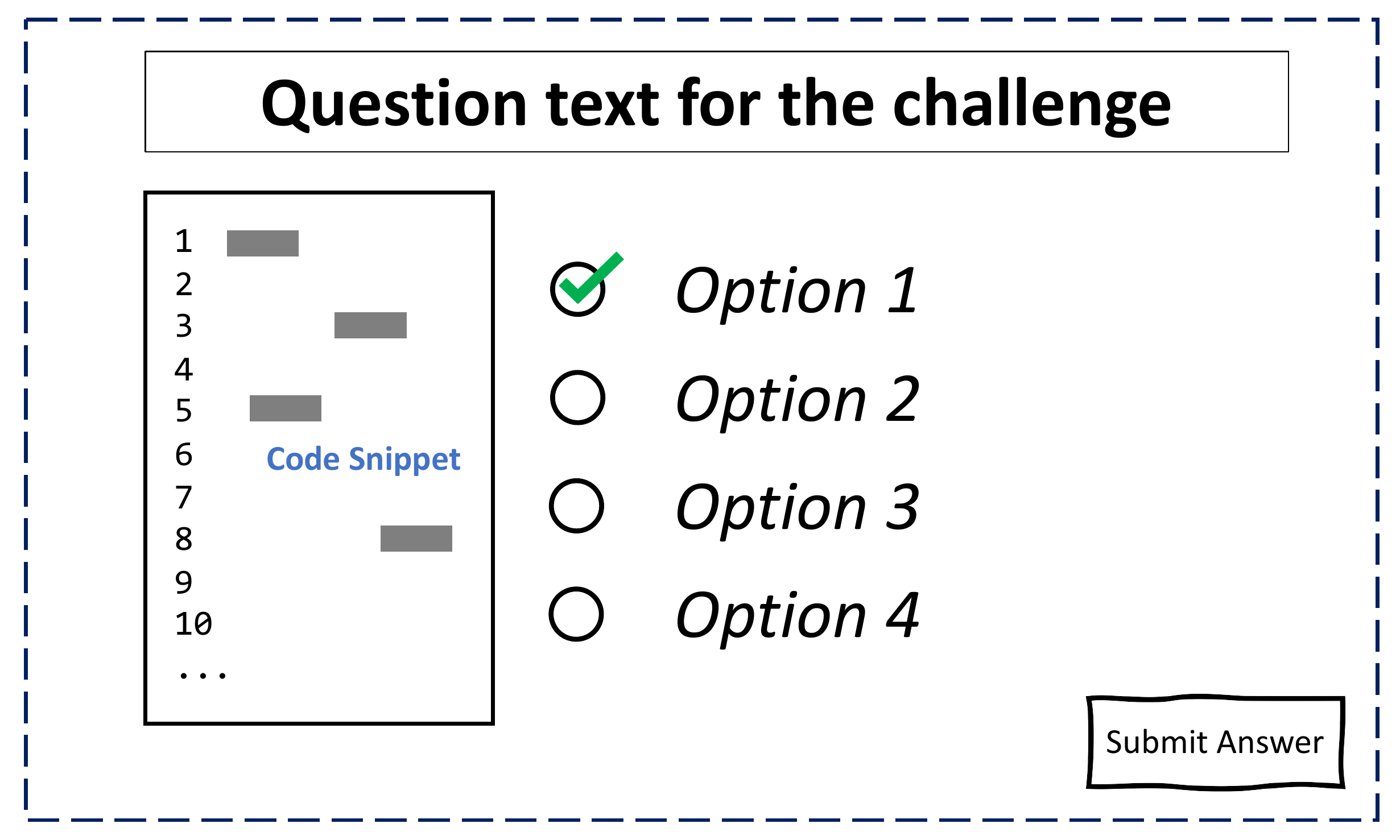}
        \vspace*{-2em}
        \caption{Code-Snippet Question}
        \label{fig:generic_csq}
    \end{minipage}
    
    \vspace*{1em}
    \begin{minipage}{.45\textwidth}
        \centering
        \includegraphics[width=.99\linewidth]{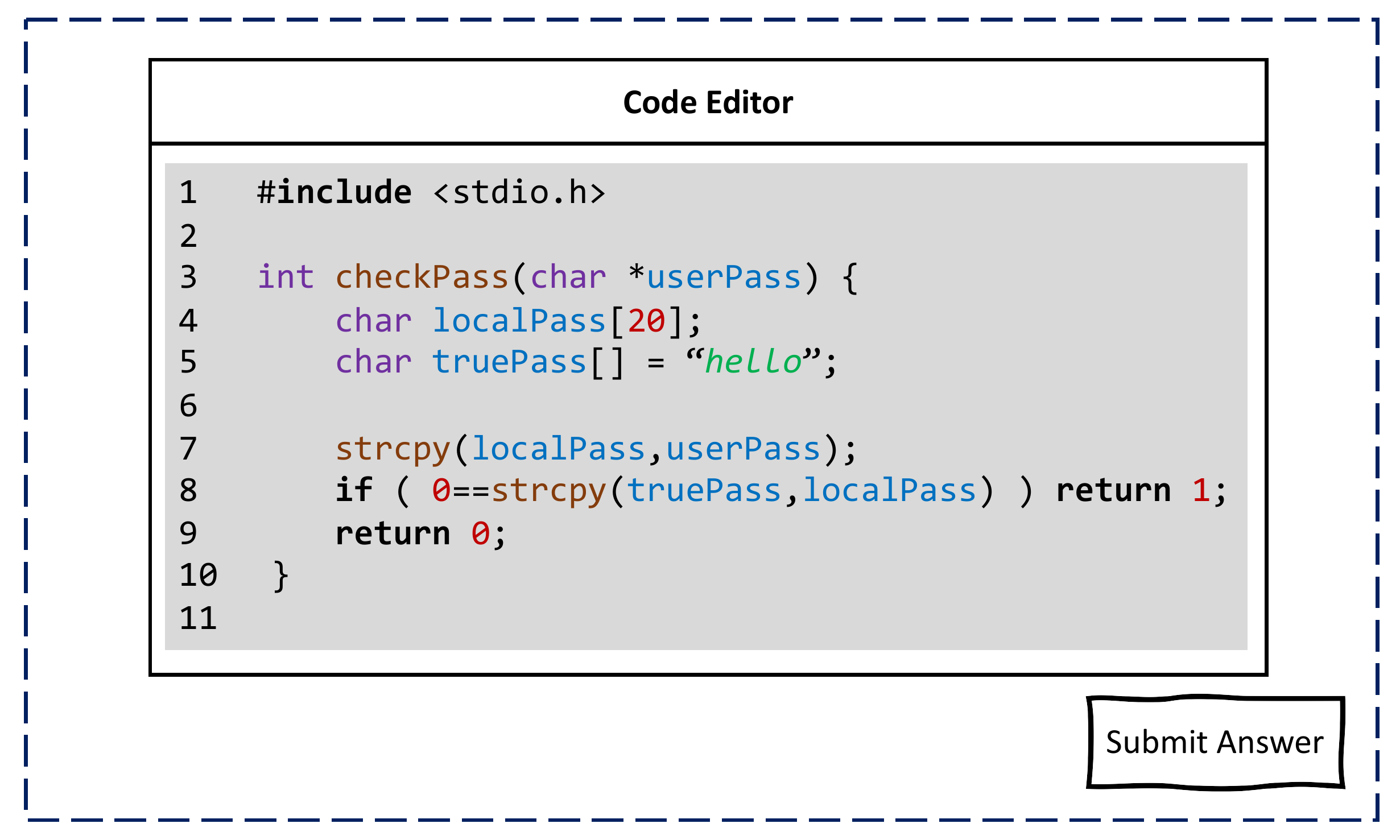}
        \vspace*{-2em}
        \caption{Code-Entry Challenge}
        \label{fig:generic_cec}
    \end{minipage}
    \begin{minipage}{.45\textwidth}
        \centering
        \includegraphics[width=.99\linewidth]{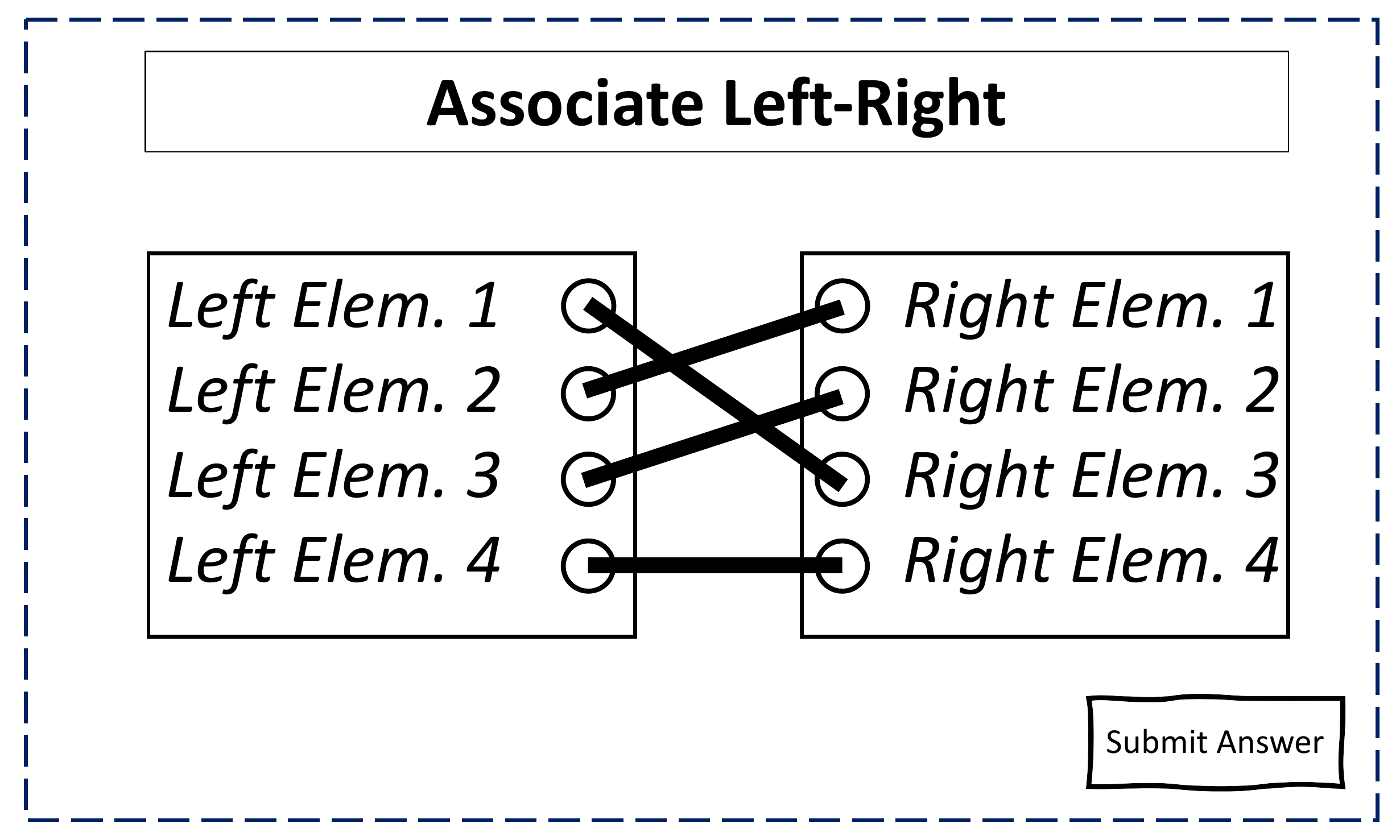}
        \vspace*{-2em}
        \caption{Associate Left-Right}
        \label{fig:generic_alr}
    \end{minipage}
\end{figure}

Figures~\ref{fig:generic_scq}-\ref{fig:generic_alr} show mock-up sketches of possible implementations on how to create a defensive CTF challenge based on the six challenge types.
Each challenge contains a guiding question, an area where the player can interact with the challenge and a \textit{submit} button to submit the results to the backend and trigger the logic stage.

\begin{table}[http]
\vspace*{1em}
\centering
\renewcommand{\arraystretch}{0.8}
{
  \footnotesize 
  \begin{tabular}{|p{4cm}|p{7.2cm}|}
    \hline
      \multicolumn{1}{|c|}{\textbf{Challenge Type}}  &
      \multicolumn{1}{c|}{\textbf{Description}}      \\
    \hline
      Single Choice Question         &
      Select a single correct answer \\
    \hline
      Multiple Choice Question        &
      Select multiple correct answers \\
    \hline
      Text Entry Challenge             &
      Type the answer to the question  \\
    \hline
      Code Snippet Challenge                           &
      Identify lines or expressions in a code snippet  \\
    \hline
      Code Entry Challenge                             &
      Write or adapt code to eliminate vulnerabilities \\
    \hline
      Associate Left-Right                             &
      Associate elements in left-list to those in right list \\
    \hline
  \end{tabular}
}
  \vspace*{1em}
  \caption[caption]{Description of the derived challenge types}
  \label{tab:question:type}
\end{table}

\subsection{Observations}
In this work, we designed defensive challenges for CTF events, which aim to raise secure coding awareness of software developers in the industry.
Code-Entry Challenges were found to be among the most popular choice, while Text-Entry Questions among the least popular.
Both the initial CS and the CT were updated as a result of the interviews with security experts.
Interestingly, the informal discussions with the security experts did not result in CTs based on comics~\cite{2019_barela}.
Another interesting observation is that all the security experts considered "simple" game types, i.e., no discussions took place on advanced challenge types based e.g., on Virtual Reality or Role-Playing-Games.
This fact is likely related to the particular nature of the topic and deployment environment (industry).
As such, challenge types that are more simple and traditional have been selected (e.g., Single-Choice Questions and Multiple-Choice Questions).
One unexpected challenge type was the Code-Entry Challenge.
Due to its complex nature, this type of challenge requires more investigation to understand how to create a challenge based on this type effectively.

\subsection{Threats to Validity}
The work hereby presented is based on the knowledge and know-how obtained through interactive discussions and surveys from a group of $20$ security experts from the industry.
A possible threat to our conclusions is the limited number of participants and their company background.

Although the authors found previous work on defensive challenges for Capture-the-Flag events, they were not focused on secure coding, software developers, and the industry.
Nevertheless, since the authors did not perform a systematic literature review, it might be that some challenge types present in scientific literature might also apply to our situation and constraints.

Another limitation of our work is that it was based only on feedback from security experts and not from players, i.e., real CTF participants.
As such, no direct feedback from the target group was used in our evaluation, especially in the preferred challenge type ranking.
The authors will address these issues in a subsequent publication.

Although the present work follows survey methodology and semi-structured interviews best practices, it lacks a systematic and academic approach.
The reason for this is that the study was conducted in an industrial setting.
However, extensive searches were conducted in scientific publication search engines to identify previous relevant work.
These findings constituted part of the initial CT and CS design.
\section{Conclusions}
\label{sec:conclusions}
Nowadays, there is an increasing demand for awareness training of software developers in the industry on secure coding.
This demand is motivated by requirements from quality standards and security security standards.
One promising new method to raise security awareness is the usage of Capture-the-Flag events. 
However, these events need to be specially designed in order to address the target audience and its requirements - software developers in the industry.

Recently the requirements that are needed for designing these games in an industrial setting have been investigated~\cite{gasiba_re19}.
However, the authors of this previous work did not provide details on the challenge types but rather requirements on the overall game.
The design of challenge types is not a trivial task, and poor quality challenges may result in inefficiencies (e.g., loss of productivity) that industrial companies are not willing to accept.
Barela~\cite{2019_barela} et al. gives one such example, which has shown that challenge types based on comics, when deployed in CTF, might not be adequate for the event and its goals.
Furthermore, the majority of the existing literature not only focuses on defensive challenges but also mostly addresses a target audience of security professionals, e.g., pen-testers.

In this work, we have addressed the design of defensive challenges for CTFs for the industry.
We have derived a challenge structure and six different challenge types.
Our work is based on semi-structured interviews with security experts and comprises a two-part survey and additional informal discussions.
Our results show that security experts prefer "\textit{traditional}" challenge types, based e.g., on Single-Choice and Multiple-Choice Questions.
We have seen that the least preferred challenge type by security experts is the Text-Entry Challenges.
Three additional challenge types have been discussed: Association Left-Right, Code-Entry Challenge, and  Code-Snippet Challenge.
The two latter types are well adapted to secure coding challenges since they use software code.

However, the unexpected new challenge type was the Code-Entry Challenge, where the player submits code to the backend, which decides if the challenge is correctly solved.
A topic that needs additional investigation is the details on how to create such a challenge type.
The results presented in this publication have been derived solely based on feedback from interviews with security experts.
Further work is required to validate the derived challenge structure and challenge types in real CTF events in an industrial setting.
In particular, the authors intend to give concrete examples of the implementation of the different derived challenge types in an upcoming publication.
This further work will allow to refine further the challenge structure and challenge types based on the feedback from the CTF players themselves.

\subsection*{Acknowledgement}
This work is financed by portuguese national funds through FCT - Fundação para a Ciência e Tecnologia, I.P., under the project FCT UIDB/04466/2020. Furthermore, the third author thanks the Instituto Universitário de Lisboa and ISTAR-IUL, for their support.

\bibliographystyle{splncs04}
\linespread{0.91}
\bibliography{bibliography}

\end{document}